\documentclass{blois}
\usepackage{epsfig}

\usepackage{bm}
\usepackage{subfigure}
\usepackage[superscript,biblabel]{cite}
\usepackage{amsfonts,amssymb,amsmath}
\usepackage{hyperref}

\def\be{\begin{equation}}
\def\ee{\end{equation}}
\def\bea{\begin{eqnarray}}
\def\eea{\end{eqnarray}}

\newcommand{\bb}{$b\bar{b}$}

\newcommand{\whl}{$WH\rightarrow \ell\nu b\bar{b}$}

\newcommand{\zhl}{$ZH\rightarrow \ell\ell b\bar{b}$}

\newcommand{\zhv}{$ZH\rightarrow \nu\bar{\nu} b\bar{b}$}

\newcommand{\vhbb}{$VH\rightarrow V b\bar{b}$}
\newcommand{\vzbb}{$VZ\rightarrow V b\bar{b}$}
\newcommand{\hbb}{$H\rightarrow b\bar{b}$}

\newcommand{\gev}{~Ge\kern -0.05em V\kern -0.1em /$c^2$}
\newcommand{\dgev}{Ge\kern -0.05em V\kern -0.1em /$c^2$}

\newcommand{\ifb}{fb$^{-1}$}

\newcommand{\met}{\ensuremath{\not\hspace*{-0.84ex}E_T\,}}

\def\citeall{\cite{Aaltonen:2013kxa,
Aaltonen:2012qt,
Aaltonen:2013ipa,
Aaltonen:2012if,
Aaltonen:2012ic,
Aaltonen:2013js,
Aaltonen:2012ii,
Aaltonen:2012id,
Aaltonen:2012ya,
Aaltonen:2012jh,
Aaltonen:2012ji,
Collaboration:2012bk,
Collaboration:2012pa,
Aaltonen:2013iia,
Abazov:2013gmz,
Abazov:2012tf,
Abazov:2012wh97,
Abazov:2013mjc,
Abazov:2012kg,
Abazov:2013alg,
Abazov:2012hv,
Abazov:2013wha,
Abazov:2013eha,
Abazov:2012zj,
Abazov:2012ee,
Abazov:2013pci
}}

\newcommand{\Photo}{}

\begin{document}
\vspace*{4cm}
\title{OVERVIEW OF HIGGS BOSON STUDIES AT THE TEVATRON}

\author{LIDIJA \v{Z}IVKOVI\'{C}}

\address{
Laboratory for High Energy Physics,
Institute of Physics Belgrade
Pregrevica 118,\\
11080 Zemun,
Serbia\\
lidiaz@fnal.gov}

\maketitle\abstracts{
The CDF and D0 experiments at the Tevatron $p\bar{p}$ Collider collected data between 2002 and 2011,
accumulating up to 10~\ifb\ of data.
During that time, an extensive search for the standard model Higgs boson was performed.
Combined results from the searches for the standard model Higgs boson with the final dataset are presented, together
with results on the Higgs boson couplings and spin and parity.}

\section{Introduction}
The standard model (SM) Higgs boson was introduced to explain electroweak symmetry 
breaking~\cite{Higgs:1964ia,Higgs:1964pj,Higgs:1966ev,Englert:1964et,Guralnik:1964eu}.
The search for the Higgs boson was a central part of the D0 and CDF Collaborations' physics program
for many years. Recently both experiments finalized their searches and published final 
results~\citeall.

\section{Search for $H\to b\bar{b}$}\label{sec:bb}

Due to the overwhelming backgrounds from multijet production, a search for the Higgs boson in the final
state with the two $b$--quarks
can not be done in the dominant production mode, $ggH$, but
 is performed in the associated production with a vector boson $V$ ($V=W,Z$).
The search is divided according to the decay of the associated vector boson: (a) \zhl, (b) \whl, and (c) \zhv.
One of the main ingredients of the search in these final states is an identification of the jets
originating from the $b$--quarks, i.e. $b$--tagging. Both the CDF and D0 collaborations
use multi--variate analysis (MVA) for $b$--tagging which improved efficiency for up to 15\%
over a cut based analysis~\cite{Freeman:2012uf,Abazov2014290}.


To validate our procedures and results 
we measure the cross sections of SM processes with similar characteristics as the Higgs boson signal.
Measured combined cross section of the \vzbb\ process is
$\sigma=(0.68\pm 0.21)\times\sigma_{SM}$.

\section{Combined results from the D0 and CDF experiments}\label{sec:Res}

\begin{figure}[htb]
\begin{center}
\includegraphics[width=0.32\textwidth]{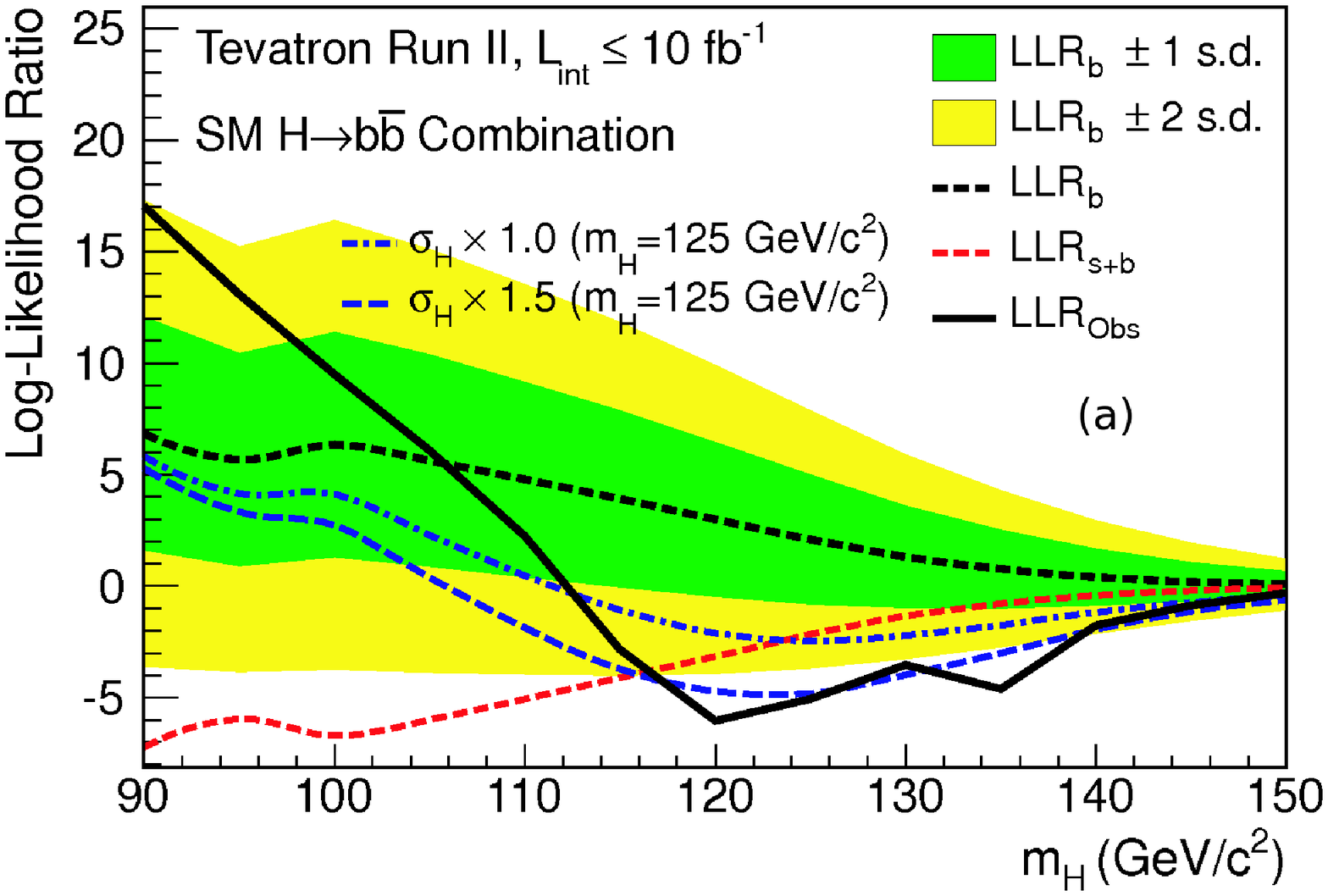}
\includegraphics[width=0.32\textwidth]{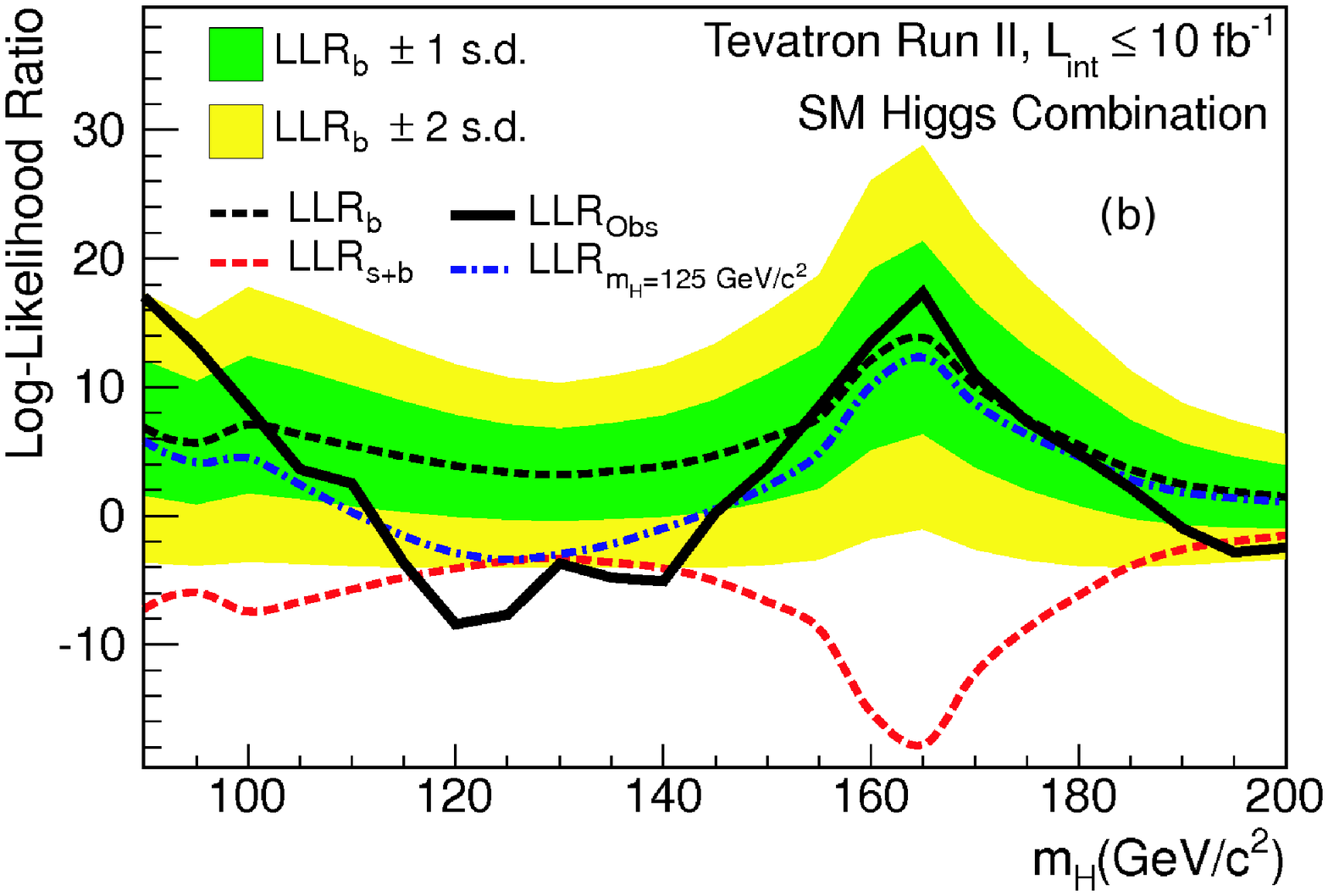}
\includegraphics[width=0.32\textwidth]{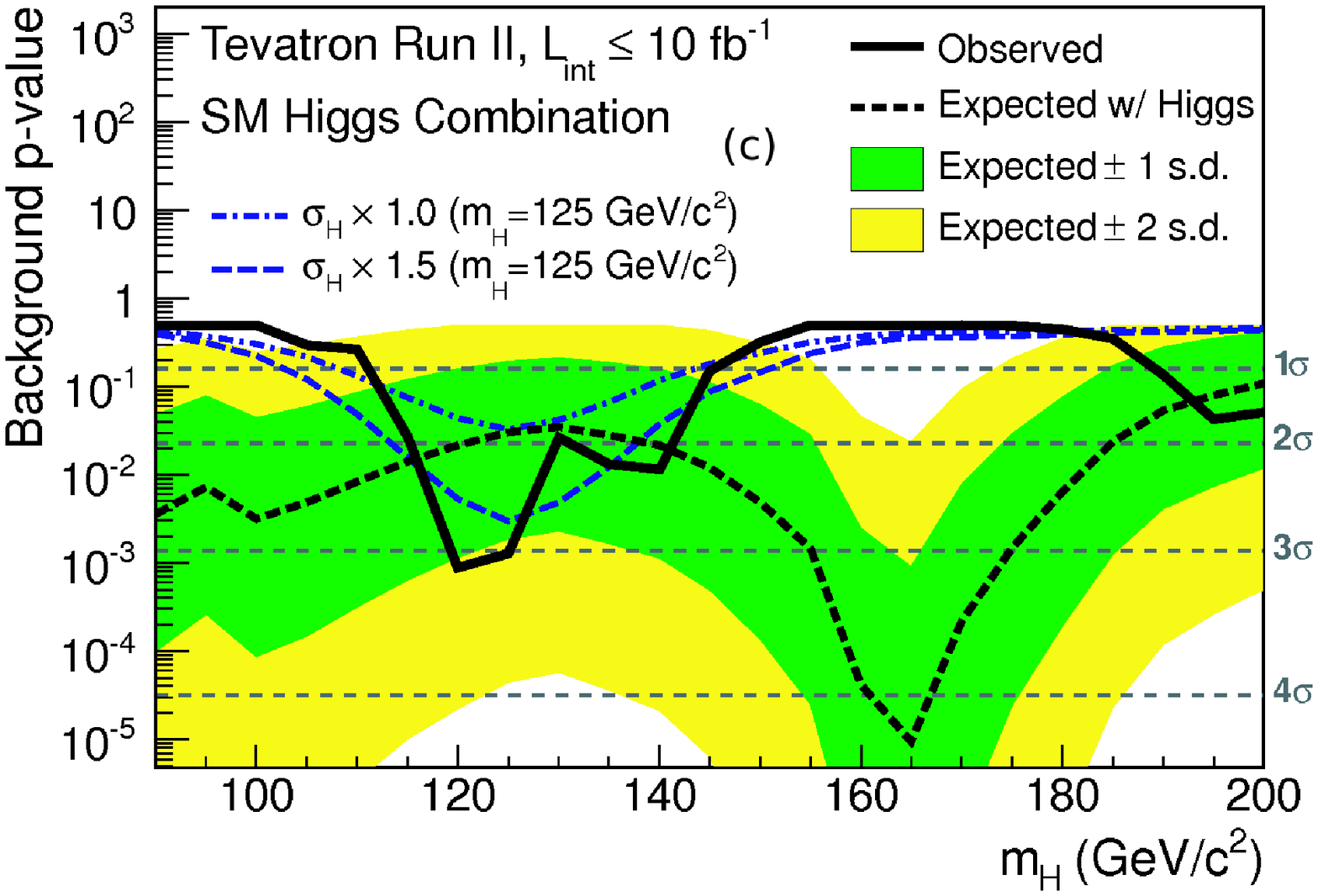}
\end{center}
\caption{ The LLR distribution for the (a) combined \vhbb\ channels, and (b) all combined channels.
(c) $p$--value for all combined channels. \label{fig:llr_p}}
\end{figure}

To estimate the sensitivity of the search at the Tevatron experiments we use the
log--likelihood ratio (LLR) test statistic for
the signal--plus--background ($s+b$) and background--only ($b$) hypotheses, defined as
${\rm LLR} = -2\ln(L_{s+b}/L_{b})$, and $L_{s+b(b)}$ is the likelihood function for the
$s+b$($b$)
hypothesis. Figure~\ref{fig:llr_p}(a) shows the LLR for the combined \vhbb\ channels from both experiments,
where data shows broad excess consistent with the dijet mass resolution.
For interpretation see Ref.~\citen{Aaltonen:2013kxa}.
We also measured the cross section of the \vhbb\ process and obtained
$(\sigma_{WH}+\sigma_{ZH})\times{\cal B}(H\to b\bar{b})=0.19^{+0.08}_{-0.09}$~pb for $M_H=125$~GeV, to be compared 
to the SM value of $0.12\pm 0.01$~pb.

When all search channels~\footnote{All search channels include \bb, $W^+W^-$, $\tau^+\tau^-$ and 
$\gamma\gamma$ decays with all production modes.}
 are combined the Tevatron experiments exclude (expect to exclude)
the Higgs boson with a mass 
90--109 (90--120) GeV and 149--182 (140--184) GeV  at the 95\% C.L.
Figure~\ref{fig:llr_p}(b) shows LLR from all search channels where an excess in data is
consistent with the assumption of the presence of the Higgs
boson with a $M_H=125$~GeV and a cross section of $\sim1.5(\pm0.6)\times\sigma_{SM}$.
To quantify this excess, 
we present in Fig.~\ref{fig:llr_p}(c)  local $p$--values for the background hypothesis,
which provide information about the
consistency with the observed data. 
We obtain that the background is inconsistent with the data at the level of 3 standard deviations (s.d.) 
for
$M_H=125$~GeV.
We define signal strength $R=\sigma/\sigma_{SM}$, and we find that our excess is consistent 
 with the presence of a SM Higgs
 boson
with $M_H=125$~GeV within one~s.d.\ with
 $R=1.44^{+0.59}_{-0.56}$. If only \hbb\ is considered, $R=1.59^{+0.69}_{-0.72}$.

\section{Couplings}\label{sec:Coup}

\begin{figure}[htb]
\begin{center}
\includegraphics[width=0.3\textwidth]{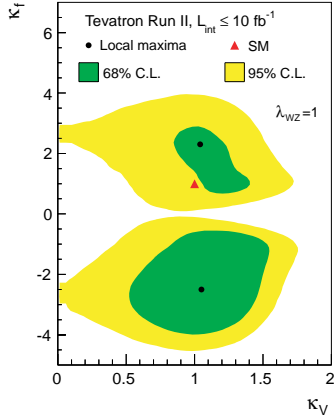}
\end{center}
\vspace*{-1cm}
\caption{
Two--dimensional constraints in the
$(\kappa_W,\kappa_Z)$ plane, for the combined Tevatron searches for a SM like
Higgs boson with mass 125 GeV.
\label{fig:coup}}
\end{figure}

After the discovery of a Higgs boson by the LHC experiments~\cite{Aad:2012tfa,Chatrchyan:2012ufa} and evidence in the
\bb\ final state at the Tevatron experiments~\cite{Aaltonen:2012qt}, it is important to precisely
measure its properties.
Any significant deviation may point to a non--SM nature of the newly discovered particle.
Since at the Tevatron experiments many production and decay modes are possible,
we assumed a
simplified model, SM--like with the following: 
(i) Hff couplings are scaled together by $\kappa_f$;
(ii) HWW coupling is scaled by $\kappa_W$;
(iii) HZZ coupling is scaled by $\kappa_Z$.
For some studies, we scale the HWW
and HZZ couplings by $\kappa_W=\kappa_Z=\kappa_V$.
Then SM is recovered if $\kappa_f=\kappa_W=\kappa_Z=1$.
Coupling results measurements from the Tevatron experiments are consistent with the SM:
(i)~assuming $\kappa_W=\kappa_Z=1$, we find $\kappa_f=-2.64^{+1.59}_{-1.30}$;
(ii)~assuming $\kappa_f=\kappa_{W(Z)}=1$, we find $\kappa_W=-1.27^{+0.46}_{-0.29}$
and $\kappa_Z=\pm 1.05^{+0.45}_{-0.55}$, and if they are varied together $(\kappa_W,\kappa_Z)=(1.25,\pm 0.90)$;
(iii)~to test custodial symmetry we study the ratio $\lambda_{WZ}=\kappa_W/\kappa_Z$, and measure
$\lambda_{WZ}=1.24^{+2.34}_{-0.42}$;
(iv)~assuming that custodial symmetry holds, $\lambda_{WZ}=1$, and allowing both $\kappa_V$ and 
$\kappa_f$ to vary, we find $(\kappa_V,\kappa_f)=(1.05, -2.40)$ and $(\kappa_V,\kappa_f)=(1.05, 2.30)$ (see Fig.~\ref{fig:coup}).

%

\section{Spin and Parity}\label{sec:JP}

\begin{figure}[htb]
\begin{center}
\includegraphics[width=0.3\textwidth]{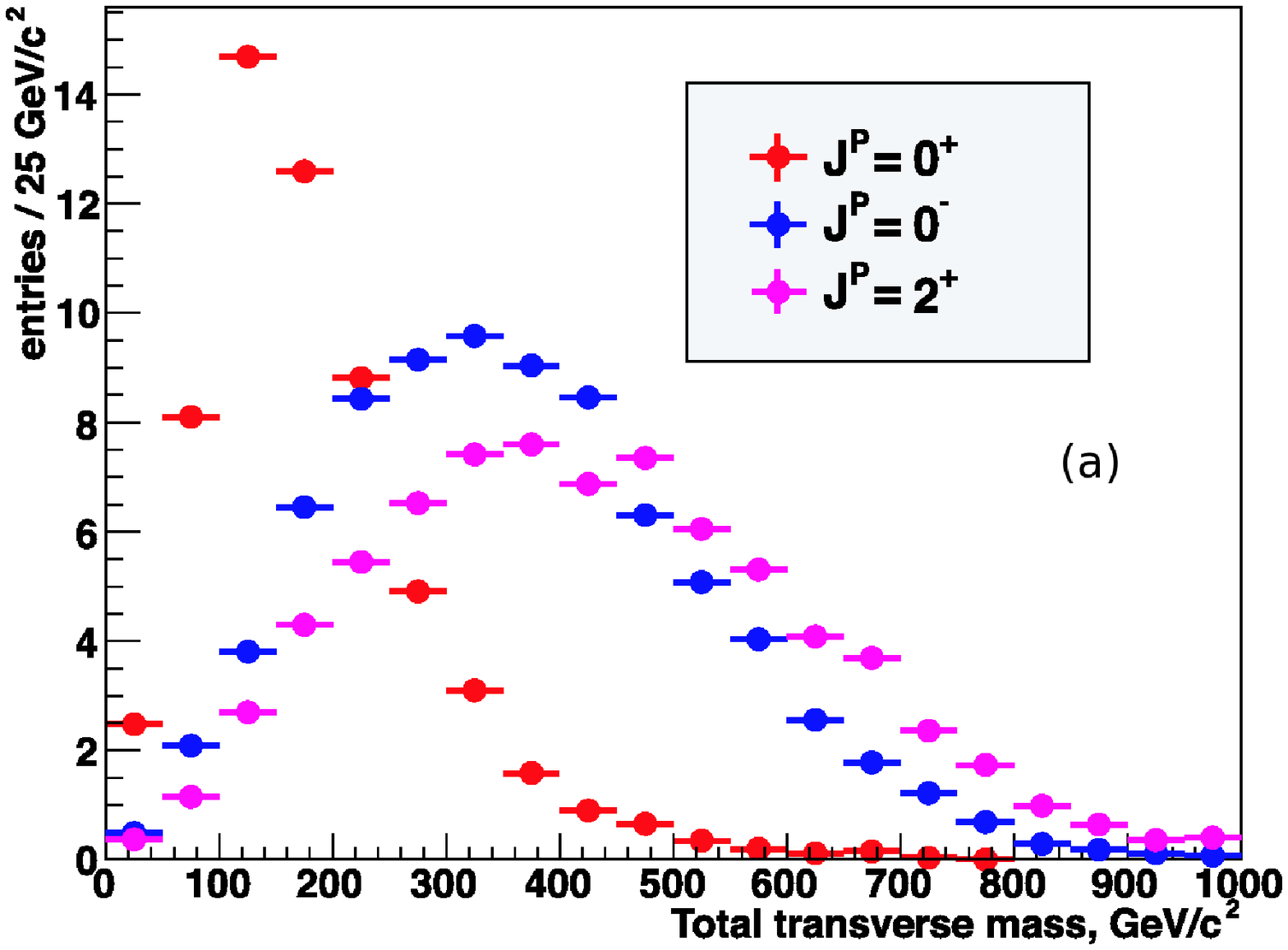}
\includegraphics[width=0.34\textwidth]{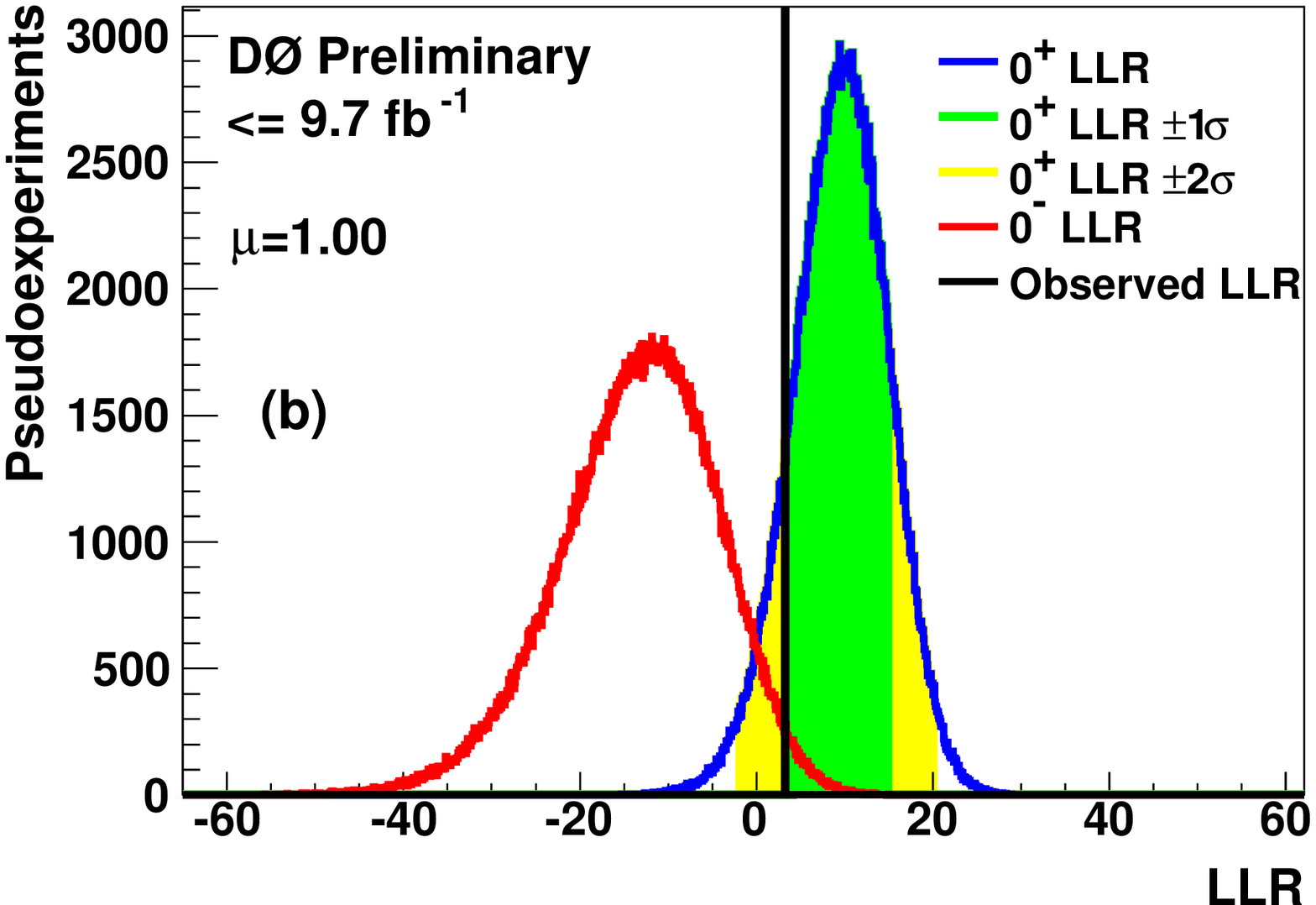}
\includegraphics[width=0.34\textwidth]{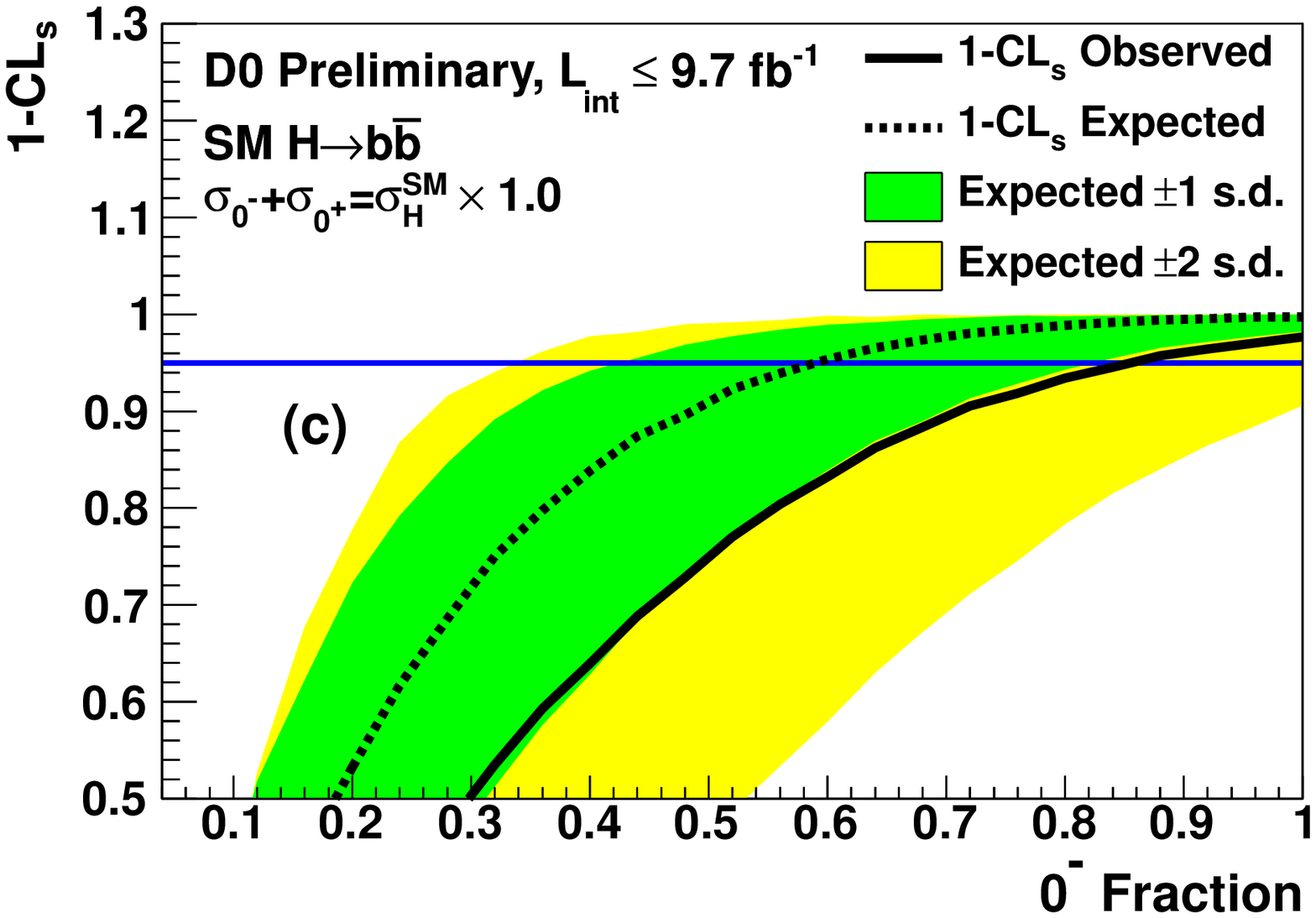}
\end{center}
\caption{(a) The transverse mass of the $VX$ system in the $WX\to\ell\nu b\bar{b}$ final state. 
(b) LLR distributions comparing the $J^P=0^+$ and the $J^P=0^-$ hypotheses for the combination of three \vhbb\ final states.
(c) $1-CL_s$ as a function of the $J^P=0^-$ signal fraction.
\label{fig:spin}}
\end{figure}

The SM predicts that Higgs boson is a scalar with a positive parity ($J^P=0^+$). 
The D0 experiment studied the spin and parity of the particle decaying to a pair of $b$--quarks,
produced in association with a vector boson,  
with assumptions that a new particle is pseudoscalar, $J^P=0^-$, or that it is graviton--like
particle described with Randall--Sundrum model, $J^P=2^+$,
following specific models described in Ref.~\citen{Ellis:2012xd}.
Associated production of a new particle $X$ and vector boson at the Tevatron is sensitive to the
different kinematics of the various $J^P$ combinations, which 
 give different $VX$ mass distributions for the different $J^P$ choices.
We use the invariant  mass for the \zhl\ channel and the transverse mass of 
the $\ell\met b\bar{b}$ and
$\met b\bar{b}$ for the \whl\ and \zhv\ channels (see Fig.~\ref{fig:spin}(a)).
To achieve better sensitivity, we divide data into samples with high and low purity based on 
dijet invariant mass, $m_{b\bar{b}}$,  in the 
$ZX\to\ell\ell b\bar{b}$ and
$VX\to\met b\bar{b}$ final states, and
MVA output used in SM Higgs boson search in the $WX\to\ell\nu b\bar{b}$ final state.

Results are combined
using the $CL_s$ method with a negative log--likelihood ratio (LLR) test statistic
$\rm{LLR}=−2\ln(L_{H_1}/L_{H_0})$, where $H_i$ is test or null hypothesis.
The $CL_s$ is then $CL_s=CL_{H_1}/CL_{H_0}$.
In the context of the models studied, 
we exclude $J^P=0^-$ at the 97.9\% C.L.\ (see Fig.~\ref{fig:spin}(b))
and $J^P=2^+$ at the 99.2\% C.L.
If we assume that both  $J^P=0^-$($J^P=2^+$) and $J^P=0^+$ are mixed in data, and
define fraction of the non--SM signal X to be
$f_X=\sigma_X/(\sigma_X+\sigma_{0^+})$, we exclude
 $f_{0^-(2^+)}>0.85(0.71)$ at 95\% C.L.\ (see Fig.~\ref{fig:spin}(c)).

\section{Summary}

In summary, we presented combined results from the D0 and CDF Collaborations in the search for the SM Higgs boson.
We report an evidence of a Higgs boson when all channels are combined.
In addition, we measured various Higgs boson couplings, and found them to be consistent with the SM.
We exclude two models of Higgs bosons
with exotic spin and parity.

\section*{Acknowledgements}

We thank the Fermilab staff and technical staffs of
the participating institutions for their vital contributions.
We acknowledge support from the DOE and NSF
(USA), ARC (Australia), CNPq, FAPERJ, FAPESP
and FUNDUNESP (Brazil), NSERC (Canada), NSC,
CAS and CNSF (China), Colciencias (Colombia), MSMT
and GACR (Czech Republic), the Academy of Finland,
CEA and CNRS/IN2P3 (France), BMBF and DFG (Germany),
DAE and DST (India), SFI (Ireland), INFN
(Italy), MEXT (Japan), the KoreanWorld Class University
Program and NRF (Korea), CONACyT (Mexico),
FOM (Netherlands), MON, NRC KI and RFBR (Russia),
the Slovak R\&D Agency, the Ministerio de Ciencia
e InnovaciÂ´on, and Programa Consolider--Ingenio 2010
(Spain), The Swedish Research Council (Sweden), SNSF
(Switzerland), STFC and the Royal Society (United
Kingdom), the A.P. Sloan Foundation (USA), and the
EU community Marie Curie Fellowship contract 302103.

The author is also supported by Serbian Ministry of Education, Science and  Technological development project 171004.

\section*{References}

\bibliography{higgs}

\end{document}

